\documentclass[11pt, a4paper]{article}
\usepackage[utf8]{inputenc}
\usepackage{amsmath,setspace,geometry}
\usepackage{amsthm}
\usepackage{amsfonts}
\usepackage[shortlabels]{enumitem}
\usepackage{rotating}
\usepackage{pdflscape}
\usepackage{graphicx}
\usepackage{bbm}
\usepackage[dvipsnames]{xcolor}
\usepackage{hyperref}
\hypersetup{colorlinks=true, linkcolor= BrickRed, citecolor = BrickRed, filecolor = BrickRed, urlcolor = BrickRed, hypertexnames = true}
\usepackage[]{natbib} 
\bibpunct[:]{(}{)}{,}{a}{}{,}
\usepackage[english]{babel}
\usepackage{float}
\usepackage{adjustbox}
\usepackage{caption}
\usepackage{subcaption}
\usepackage{booktabs}
\usepackage{pdfpages}
\usepackage{threeparttable}
\usepackage{lscape}
\usepackage{bm}
\title{Challenges in Statistically Rejecting the Perfect Competition Hypothesis Using Imperfect Competition Data}
\author{Yuri Matsumura\thanks{Department of Economics, Rice University. Email: \href{mailto:}{yuri.matsumura23@gmail.com}} \and Suguru Otani \thanks{\href{mailto:}{suguru.otani@e.u-tokyo.ac.jp}, Market Design Center, University of Tokyo
}
\thanks{
The title of the earlier version is ``Finite Sample Performance of a Conduct Parameter Test in Homogeneous Goods Markets".
We thank Jeremy Fox, Isabelle Perrigne, and Yuya Shimizu for their valuable advice. This research did not receive any specific grant from funding agencies in the public, commercial, or not-for-profit sectors. }}

\date{
First version: November 14, 2023\\
Current version: \today
}

\begin{document}

\maketitle
\begin{abstract}
    We theoretically prove why statistically rejecting the null hypothesis of perfect competition is challenging, known as a common problem in the literature. We also assess the finite sample performance of the conduct parameter test in homogeneous goods markets, showing that statistical power increases with the number of markets, a larger conduct parameter, and a stronger demand rotation instrument. However, even with a moderate number of markets and five firms, rejecting the null hypothesis of perfect competition remains difficult, irrespective of instrument strength or the use of optimal instruments. Our findings suggest that empirical results failing to reject perfect competition are due to the limited number of markets rather than methodological shortcomings.
\vspace{0.1in}

\noindent\textbf{Keywords:} Conduct parameters, Homogeneous goods market, Monte Carlo simulation, Statistical power analysis
\vspace{0in}
\newline
\noindent\textbf{JEL Codes:} C5, C12, L1

\bigskip
\end{abstract}

\section{Introduction}
Measuring competitiveness is important in the empirical industrial organization literature.
A conduct parameter is considered to be a useful measure of competitiveness. 
However, the parameter cannot be directly measured from data because data generally lack information about marginal costs.
Therefore, researchers endeavor to learn conduct parameters.

Researchers estimate and test structural models to understand firm conduct in both homogeneous and differentiated goods markets \citep{bresnahan1982oligopoly,nevoIdentificationOligopolySolution1998, magnolfi2022comparison, duarte2023testing}.
This paper focuses on homogeneous goods markets and testing based on the conduct parameter model.
As noted by several papers \citep{genesove1998testing, steen1999testing, shaffer1993test}, a well-known issue in testing is that the null hypothesis of perfect competition cannot be rejected when the researcher uses imperfect competition data.
Table \ref{tab:conduct_parameter_testing_literature} includes the papers that face the problem.
Although the problem is widely applied, there is a notable lack of formal proof explaining the problem.
This gap in the literature hinders our ability to fully understand the aforementioned comparisons.

In response, we provide a theoretical explanation for the problem.
Subsequently, we construct optimal instruments that make it possible for us to simulate the best scenario in which researchers can obtain the most efficient moment conditions.
Finally, we investigate the finite sample performance of the conduct parameter test in homogeneous goods markets.
We analyze statistical power by varying the number of markets, the number of firms, and the strength of demand rotation instruments, all under the null hypothesis of perfect competition.

Our simulation results indicate that statistical power increases with a greater number of markets, a larger conduct parameter, and stronger demand rotation instruments. 
However, even with a moderate number of markets (e.g., 1000) and five firms, achieving an 80\% rejection frequency ($1-\beta=0.8$, where $\beta$ denotes the probability of a Type II error) remains elusive, regardless of instrument strength or the use of optimal instruments. 
While optimal instruments improve rejection probability in large samples, they do not alter the fundamental findings.

These results, which align with the theoretical statistical power formula, underscore the inherent difficulty of testing perfect competition.
This difficulty arises primarily from the limited number of markets rather than from methodological deficiencies.

Our results and accompanying code provide a valuable resource for applied researchers investigating assumptions about firm conduct in homogeneous goods markets, whether considering perfect competition, Cournot competition, or perfect collusion.

\begin{table}[t!]
    \begin{adjustbox}{width=1\textwidth,center}
        \begin{tabular}{|l|c|c|c|l|}
        \hline
        \textbf{Paper} & \textbf{Industry} & \textbf{Markets} & \textbf{Estimated $\theta$} & \textbf{Test results} \\ \hline
        \cite{genesove1998testing} & Sugar & 97 & 0.04 (0.1)& Cannot reject PC \\ 
        \cite{clay2003further} & Whiskey & 38 & 0.12-0.35 (0.09) & Results vary \\ 
        \cite{steen1999testing} & Salmon & 48 & 0.02 & Cannot reject PC \\ 
        \cite{shaffer1993test} & Banking & 25 & 0.00 & Cannot reject PC \\
        \cite{bettendorf2000incomplete} & Coffee bean & 60 & 0.02-0.17 & Results vary  \\
        \cite{wolfram1999measuring} & Electricity & 25,639 & 0.01(0.05) & Cannot reject PC \\ 
        \cite{kim2006biases} & Electricity & 21,104 & 0.12-0.23 (0.07) & Reject PC \\ 
        \cite{puller2007pricing} & Electricity & 163 - 573 & (0.2) & Cannot reject Cournot \\ 
         \hline
        \end{tabular}
    \end{adjustbox}
    \caption{Related studies on conduct parameter testing in homogenous goods industries}
    \label{tab:conduct_parameter_testing_literature}
    \footnotesize
    \vspace{2mm}
    Note: The number of parenthesis shows a directly measured conduct parameter. $\theta$ means an industry-average conduct parameter. PC means a perfect competition null hypothesis. \cite{puller2007pricing} estimate firm-level conducts from a directly-measured conduct parameter. For comparison, we convert firm-level conduct into the industry-average conduct by dividing the number of firms, i.e., five.
\end{table}

\subsection{Related Literature}

The conduct parameter model has been used in measuring competitiveness, especially in homogeneous product markets.
The conduct parameters for the linear model are identified by \citet{bresnahan1982oligopoly}, while \cite{matsumura2023resolving} further explore the condition for the identification.
\citet{lau1982identifying} considers the identification of the conduct parameter in a more general model.
Estimation accuracy can be enhanced by incorporating equilibrium existence conditions in the log-linear model \citep{matsumura2024test}. 

Table \ref{tab:conduct_parameter_testing_literature} summarizes the literature on testing the conduct parameter.
Conduct parameter testing has been pursued by \cite{genesove1998testing}, who compare estimates from the sugar industry with direct measures of market power.\footnote{\cite{genesove1998testing} made errors in obtaining predicted interaction terms of rotation demand instruments and endogenous quantities in the first-stage regression. See Section A.3 of \cite{matsumura2023resolving}.} 
When market power is approximately 0.1 and the number of markets is fewer than 100, the null hypothesis of perfect competition cannot be rejected.
\cite{clay2003further} investigate the robustness of specifications in 38 whiskey markets and find that estimation is sensitive to model specification. 
Similarly, \cite{steen1999testing} examine 48 markets in the French salmon industry, \cite{shaffer1993test} analyzes 25 markets in the Canadian banking industry, \cite{bettendorf2000incomplete} study 60 monthly markets in the Dutch coffee bean industry. 
These studies also fail to reject the null hypothesis of perfect competition, raising concerns about the methodology itself \citep{shafferMarketPowerCompetition2017}.

Some papers compare estimated conduct parameters with the competitiveness directly measured through price-cost markups derived from observed prices and marginal cost data.
\cite{wolfram1999measuring} analyzes the British electricity industry using 25,639 samples and finds that the directly measured conduct parameter is approximately 0.05, while the estimated conduct parameter is 0.01, leading to an inability to reject the null hypothesis of perfect competition. 
\cite{kim2006biases} examine the California electricity market using 21,104 samples and find that Bresnahan's technique overstates marginal costs on average, making it more likely to reject the null hypothesis of perfect competition. 
In contrast, \cite{puller2007pricing}, using between 163 and 573 samples, is unable to reject the null hypothesis of Cournot competition for the same industry. 
The robustness of the data and methodologies has been extensively tested in previous studies on market power in the California electricity market, including those by \cite{borenstein2002measuring}, \cite{wolak2003measuring}, and \cite{orea2018estimating}.

\section{Model}
Consider data with $T$ markets with homogeneous products.
Assume that there are $N$ firms in each market.
Let $t = 1,\ldots, T$ be the index for markets.
Then, we obtain a supply equation:
\begin{align}
     P_{t} = -\theta\frac{\partial P_{t}(Q_{t})}{\partial Q_{t}}Q_{t} + MC_{t}(Q_{t}),\label{eq:supply_equation}
\end{align}
where $Q_{t}$ is the aggregate quantity, $P_{t}(Q_{t})$ is the inverse demand function, $MC_{t}(Q_{t})$ is the marginal cost function, and $\theta\in[0,1]$ is  the conduct parameter. 
The equation nests perfect competition ($\theta=0$), Cournot competition ($\theta=1/N$), and perfect collusion ($\theta=1$) (See \cite{bresnahan1982oligopoly}). 

Consider an econometric model integrating the above.
Assume that the demand and marginal cost functions are written as: 
\begin{align}
    P_{t} = f(Q_{t}, Y_{t}, \varepsilon^{d}_{t}, \alpha), \label{eq:demand}\\
    MC_{t} = g(Q_{t}, W_{t}, \varepsilon^{c}_{t}, \gamma),\label{eq:marginal_cost}
\end{align}
where $Y_{t}$ and $W_{t}$ are vectors of exogenous variables, $\varepsilon^{d}_{t}$ and $\varepsilon^{c}_{t}$ are error terms, and $\alpha$ and $\gamma$ are vectors of parameters.
Additionally, we have demand- and supply-side instruments, $Z^{d}_{t}$ and $Z^{c}_{t}$, and assume that the error terms satisfy the mean independence conditions, $E[\varepsilon^{d}_{t}\mid Y_{t}, Z^{d}_{t}] = E[\varepsilon^{c}_{t} \mid W_{t}, Z^{c}_{t}] =0$.

\subsection{Linear demand and cost}
For illustration, we assume that linear inverse demand and marginal cost functions are specified as:
\begin{align}
    P_{t} &= \alpha_0 - (\alpha_1 + \alpha_2Z^{R}_{t})Q_{t} + \alpha_3 Y_{t} + \varepsilon^{d}_{t},\label{eq:linear_demand}\\
    MC_{t} &= \gamma_0  + \gamma_1 Q_{t} + \gamma_2 W_{t} + \gamma_3 R_{t} + \varepsilon^{c}_{t},\label{eq:linear_{t}arginal_cost}
\end{align}
where $W_{t}$ and $R_{t}$ are exogenous cost shifters and $Z^{R}_{t}$ is Bresnahan's demand rotation instrument. 
The supply equation is written as:
\begin{align}
    P_{t} 
    &= \gamma_0 + \theta (\alpha_1 + \alpha_2 Z^{R}_{t})Q_{t} + \gamma_1 Q_{t} + \gamma_2 W_{t} + \gamma_3 R_{t} +\varepsilon^c_{t}.\label{eq:linear_supply_equation}\end{align}
By substituting \eqref{eq:linear_demand} with \eqref{eq:linear_supply_equation} and solving it for $P_{t}$, we obtain the aggregate quantity $Q_{t}$ based on the parameters and exogenous variables as follows:
\begin{align}
    Q_{t} =  \frac{\alpha_0 + \alpha_3 Y_{t} - \gamma_0 - \gamma_2 W_{t} - \gamma_3 R_{t} + \varepsilon^{d}_{t} - \varepsilon^{c}_{t}}{(1 + \theta) (\alpha_1 + \alpha_2 Z^{R}_{t}) + \gamma_1}.\label{eq:quantity_linear}
\end{align}

\subsection{Statistical Power Formula in Conduct Parameter Testing}
We provide a proof explaining why statistically rejecting the null hypothesis of Cournot and perfect competition is challenging. Following Chapters 5, 7, and 9 of \cite{hansen2022econometrics}, we consider the null hypothesis $H_{0}: \theta = 0$, indicating markets are perfectly competitive, and the alternative hypothesis $H_{1}: \theta \neq 0$. 

Let $T(\theta)$ denote the t-statistic of $\theta$ and $\hat{\theta}$ be the point estimator. For simplicity, we assume that the variance of the estimator for $\theta$, $V(\theta)$, is known. Then, the standard error of $\theta$ is expressed as $SE(\theta) = \sqrt{T^{-1}V(\theta)}$. 
At a significance level of 0.05, the power $\pi(\theta)$ is given by:
\begin{align}
    \pi(\theta) &= \Pr(T(\theta)>1.96|\theta\neq 0)\nonumber\\
    &=\Pr\left(\frac{\hat{\theta}}{SE(\theta)}>1.96|\theta\neq 0\right)\nonumber\\
    &=1 - \Pr\left(\frac{\hat{\theta}}{SE(\theta)}<1.96|\theta\neq 0\right)\nonumber\\
    &=1 - \Pr\left(\frac{\sqrt{T}\hat{\theta}}{\sqrt{V(\theta)}}<1.96|\theta\neq 0\right)\nonumber\\
    &\approx 1 - \Phi\left(1.96 - \frac{\sqrt{T}\hat{\theta}}{\sqrt{V(\theta)}}\right),\label{eq:statistical_power_formula}
\end{align}
for large $T$, where $\Phi(\cdot)$ is the cumulative distribution function of the normal distribution.

This formula illustrates that the term $\sqrt{T}\hat{\theta}$ determines the statistical power under $\theta \in [0,1]$. For instance, if $\theta = 0.1$ (corresponding to symmetric ten-firm Cournot competition), the number of markets, $T$, must increase by a factor of 100 compared to the case where $\theta = 1.0$ to achieve the same power. This observation aligns with our simulation findings in Figure \ref{fg:theta_hat_power}. Notably, this relationship applies to all regression coefficients.

\subsection{Optimal instruments}\label{sec:optimal_instruments}

Our statistical power formula \eqref{eq:statistical_power_formula} for conduct parameter testing is contingent upon the variance of the estimator for $\theta$, denoted as $V(\theta)$. To minimize this variance, we introduce optimal instruments, which provide the most efficient moment conditions. The case with optimal instruments represents the best scenario in which researchers can obtain the most efficient moment conditions, as discussed in our simulations. 

We begin by introducing optimal instruments for a simultaneous equation model. Following this, we turn to supply-side estimation, since the simultaneous equation model offers no additional efficiency advantage when the error terms in the demand and supply equations are uncorrelated, as is the case in our scenario.

We define demand and supply residuals as follows.
\begin{align*}
    \varepsilon^{d}_{t}(\xi) &= P_{t} - \alpha_0 + (\alpha_1 + \alpha_2Z^{R}_{t})Q_{t} - \alpha_3 Y_{t},\\
    \varepsilon^c_{t}(\xi) &= P_{t} - \gamma_0 - \theta (\alpha_1 + \alpha_2 Z^{R}_{t})Q_{t} - \gamma_1 Q_{t} - \gamma_2 W_{t} - \gamma_3 R_{t},
\end{align*}
where $\xi=\left(\alpha_0, \alpha_1, \alpha_2, \alpha_3, \gamma_0, \gamma_1, \gamma_2, \gamma_3, \theta\right)$ is the $9\times 1$ parameter vector. 
Let $Z_{t}=(Z_{t}^{d},Z_{t}^{s})$ and $\varepsilon_{t}(\xi)=(\varepsilon_{t}^{d}(\xi),\varepsilon_{t}^{s}(\xi))'$.
We make the assumption that $\Omega=E[\varepsilon_{t}\varepsilon_{t}'|Z_{t}]$ is a constant $2\times 2$ matrix, defining the covariance structure of the demand and supply residuals.
The optimal instrument matrix of \cite{chamberlain1987asymptotic} is the $9\times 2$ matrix $g_{t}(Z_{t})=D_{t}(Z_{t})'\Omega^{-1}$
where the $2\times 9$ matrix $D_{t}(Z_{t})=E\left[\frac{\partial \varepsilon_{t}}{\partial \xi}| Z_{t}\right]$ is the conditional expectation of the derivative of the conditional moment restrictions with respect to parameters $\xi$. 

The conditional expectation $D_{t}(Z_{t})$ is difficult to compute, so most applications have considered approximations. As in \cite{reynaert2014improving}, we consider two types of approximation. 
The first approximation for $D_{t}(Z_{t})$ takes a second-order polynomial of $Z_{t}$ for the demand side instruments (first row of $D_{t}(Z_{t})$), i.e. cost shifters $W_{t}$ and $R_{t}$, and their squares and interactions, and $W_{t}$ and $R_{t}$ for the supply side instruments (second row of $D_{t}(Z_{t})$).
The second approximation for $D_{t}(Z_{t})$ implements the conditional expectation written as 
\begin{align*}
    &D_{t}(Z_{t})\\
    &=E\left[\frac{\partial \varepsilon_{t}}{\partial \xi}| Z_{t}\right]\\
    &=
    \begin{pmatrix}
    & E\left[\frac{\partial \varepsilon_{t}^{d}(\xi)}{\partial \alpha^{\prime}} \mid Z_{t}\right] & 
    E\left[\frac{\partial \varepsilon_{t}^{d}(\xi)}{\partial \gamma^{\prime}} \mid Z_{t}\right] & 
    E\left[\frac{\partial \varepsilon_{t}^{d}(\xi)}{\partial \theta} \mid Z_{t}\right]\\
    & E\left[\frac{\partial \varepsilon_{t}^{c}(\xi)}{\partial \alpha^{\prime}} \mid Z_{t}\right] & 
    E\left[\frac{\partial \varepsilon_{t}^{c}(\xi)}{\partial \gamma^{\prime}} \mid Z_{t}\right] & 
    E\left[\frac{\partial \varepsilon_{t}^{c}(\xi)}{\partial \theta} \mid Z_{t}\right]
    \end{pmatrix} \\
    &=\begin{pmatrix}
    -1 & 
    \overline{Q}_{t} & 
    Z^{R} \overline{Q}_{t} & 
    - Y_{t} &
    0 & 0 & 0 & 0 & 0 \\
    0 &- \theta \overline{Q}_{t} & -\theta Z^{R}_{t}\overline{Q}_{t} & 0 & 
    -1 &
    - \overline{Q}_{t} &
    -W_{t} &
    -R_{t} &
    -(\alpha_1 + \alpha_2 Z^{R}_{t})\overline{Q}_{t}
    \end{pmatrix},
\end{align*}
where $ \overline{Q}_{t} \equiv E[Q_{t}\mid Z_{t}]$. Following the literature, we replace $ \overline{Q}_{t}$ with the fitted values of regressions of $Q_{t}$ on the second order polynomials of $(Y_{t}, Z_{t}^{R}, W_{t}, R_{t}, H_{t}, K_{t})$ and the interactions.

When only the supply side is estimated given fixed demand parameters as in our main results, the earlier optimal instrument matrix is modified into the $5\times 1$ matrix $g_{t}(Z_{t})=D_{t}(Z_{t})'$ where 
\begin{align*}
    D_{t}(Z_{t}) &= \begin{pmatrix}
     & 
    E\left[\frac{\partial \varepsilon_{t}^{c}(\xi)}{\partial \gamma^{\prime}} \mid Z_{t}\right] & 
    E\left[\frac{\partial \varepsilon_{t}^{c}(\xi)}{\partial \theta} \mid Z_{t}\right]
    \end{pmatrix} \\
    &=\begin{pmatrix}
    -1 &
    -\overline{Q}_{t} &
    -W_{t} &
    -R_{t} &
    -(\alpha_1 + \alpha_2 Z^{R}_{t})\overline{Q}_{t}
    \end{pmatrix}.
\end{align*}
This suggests augmenting the benchmark instruments with $\overline{Q}_{t}$ and $(\alpha_1 + \alpha_2 Z^{R}{t})\overline{Q}_{t}$.
To prevent redundancy, we add $\overline{Q}_{t}$ to the benchmark model.
We verify that including $(\alpha_1 + \alpha_2 Z^{R}{t})\overline{Q}_{t}$ in addition to $\overline{Q}_{t}$ does not enhance statistical power.


\section{Simulation results}\label{sec:results}

\subsection{Simulation and estimation procedure}

We set true parameters and distributions as shown in Table \ref{tb:parameter_setting}. 
We vary the true value of $\theta$ from 0.05 (20-firms symmetric Cournot) to 1 (perfect collusion) and the strength of demand rotation instrument, $\alpha_2$, from 0.1 (weak) to 20.0 (extremely strong) which is unrealistically larger than the price coefficient level, $\alpha_1=1.0$.
For simulation, we generate 100 datasets.
We separately estimate the demand and supply equations via two-stage least squares (2SLS) estimation.
The instrumental variables for demand estimation are $Z^{d}_{t} = (1, Z^{R}_{t}, Y_{t}, H_{t}, K_{t})$ and for supply estimation are $Z^{c}_{t} = (1, Z^{R}_{t}, W_{t}, R_{t}, Y_{t})$ for a benchmark model. 
To achieve theoretical efficiency bounds, we add optimal instruments of \cite{chamberlain1987asymptotic}, used in demand estimation \citep{reynaert2014improving}. 
Optimal instruments lead to asymptotically efficient estimators, as their asymptotic variance cannot be reduced via additional orthogonality conditions.
The null hypothesis is that markets are under perfect competition, that is, $\theta=0$.
We compute the rejection
frequency as the power by using $t$-statistics at a significance level of 0.05 over 100 datasets.

\begin{table}[!htbp]
    \caption{True parameters and distributions}
    \label{tb:parameter_setting}
    \begin{center}
    \subfloat[Parameters]{
    \begin{tabular}{cr}
            \hline
            $\alpha_0$ & $10.0$  \\
            $\alpha_1$ & $1.0$  \\
            $\alpha_2$ & $\{0.1,0.5,1.0,5.0,20.0\}$ \\
            $\alpha_3$ & $1.0$  \\
            $\gamma_0$ & $1.0$ \\
            $\gamma_1$ & $1.0$  \\
            $\gamma_2$ & $1.0$ \\
            $\gamma_3$ & $1.0$\\
            $\theta$ & $\{0.05,0.1,0.2,0.33,0.5,1.0\}$ \\
            \hline
        \end{tabular}
    }
    \subfloat[Distributions]{
    \begin{tabular}{crr}
            \hline
            Demand shifter&  \\
            $Y_{t}$ & $N(0,1)$  \\
            Demand rotation instrument&   \\
            $Z^{R}_{t}$ & $N(10,1)$ \\
            Cost shifter&    \\
            $W_{t}$ & $N(3,1)$  \\
            $R_{t}$ & $N(0,1)$   \\
            $H_{t}$ & $W_{t}+N(0,1)$  \\
            $K_{t}$ & $R_{t}+N(0,1)$   \\
            Error&  &  \\
            $\varepsilon^{d}_{t}$ & $N(0,\sigma)$  \\
            $\varepsilon^{c}_{t}$ & $N(0,\sigma)$ \\
            \hline
        \end{tabular}
    }
    \end{center}
    \footnotesize
    Note: $\sigma=1.0$. $N:$ Normal distribution. $U:$ Uniform distribution.
\end{table}

\subsection{Main results}

\begin{figure}[!ht]
  \begin{center}
  \subfloat[$T=100$]{\includegraphics[width = 0.32\textwidth]
  {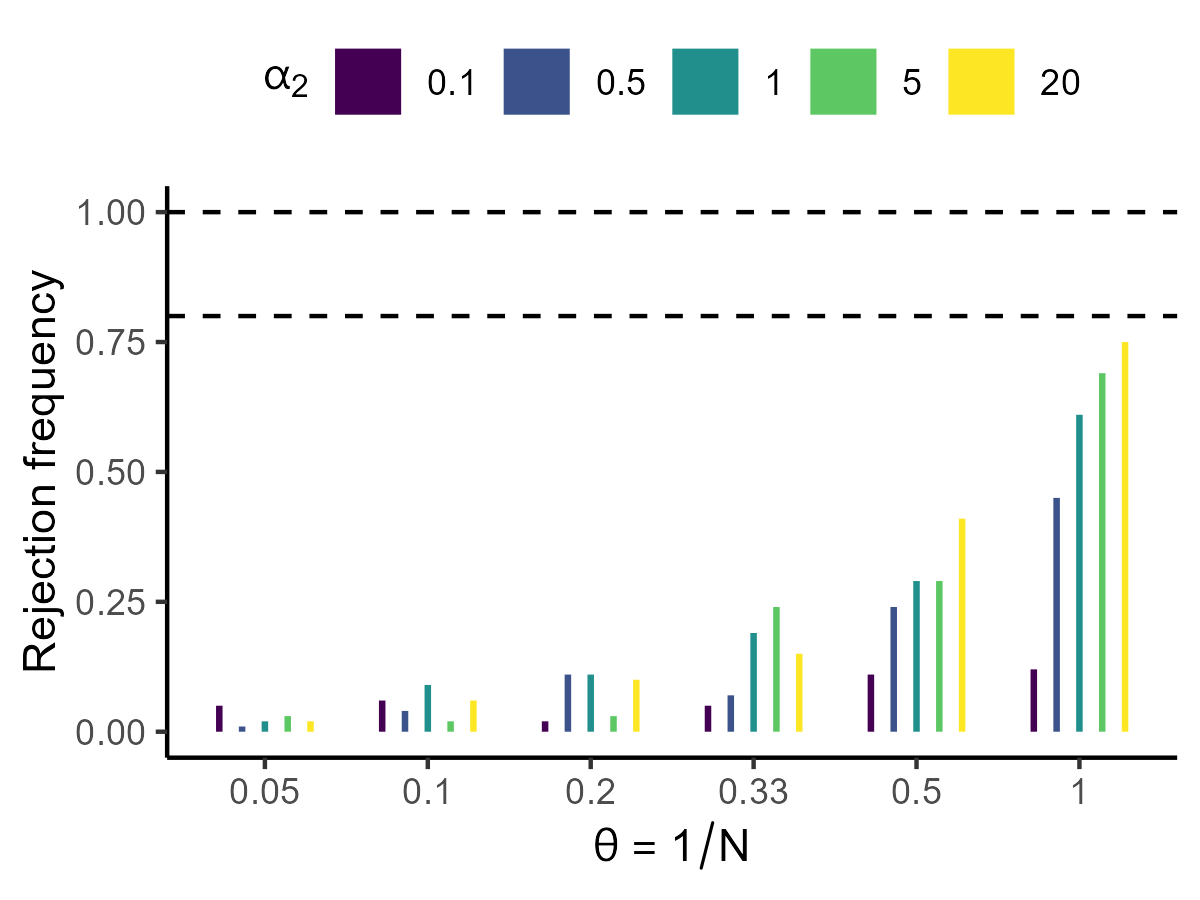}}
  \subfloat[$T=200$]{\includegraphics[width = 0.32\textwidth]
  {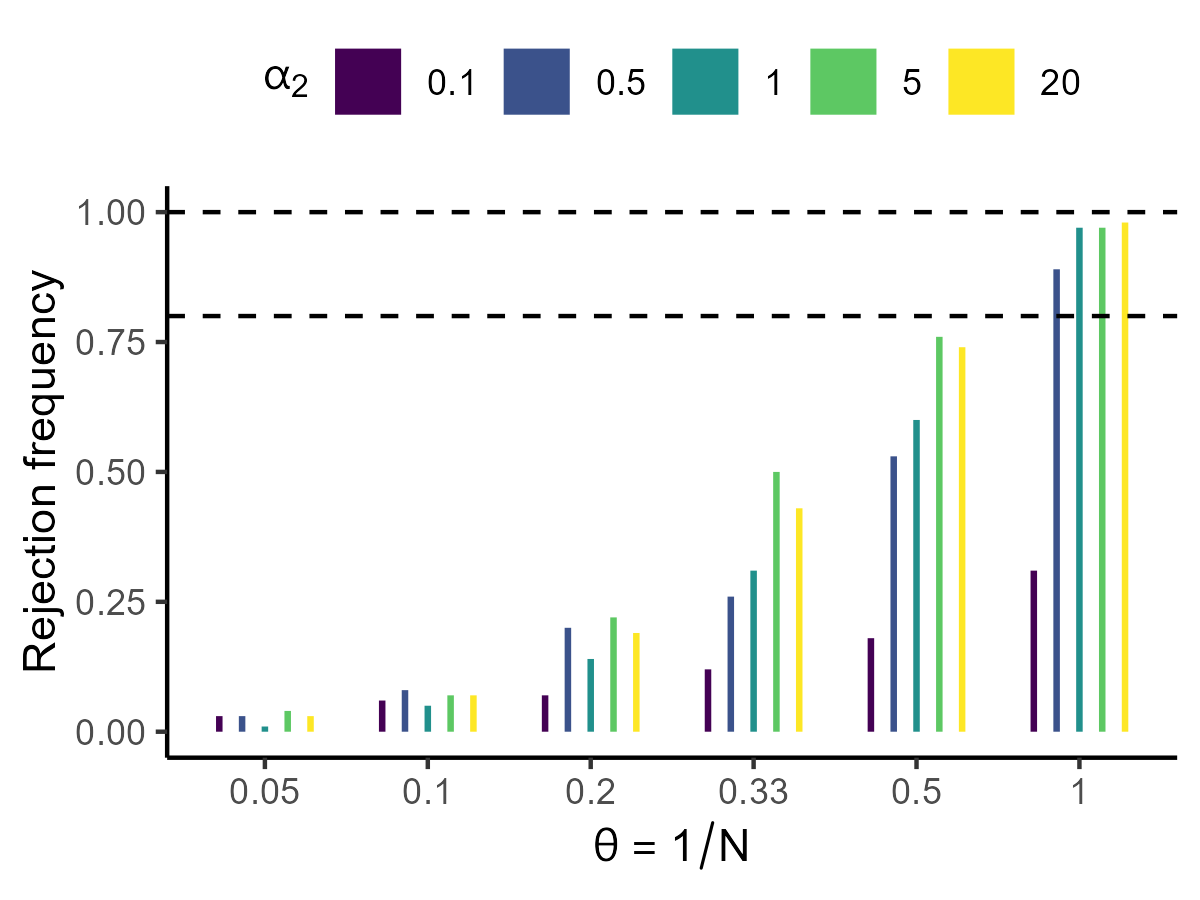}}
  \subfloat[$T=1000$]{\includegraphics[width = 0.32\textwidth]
  {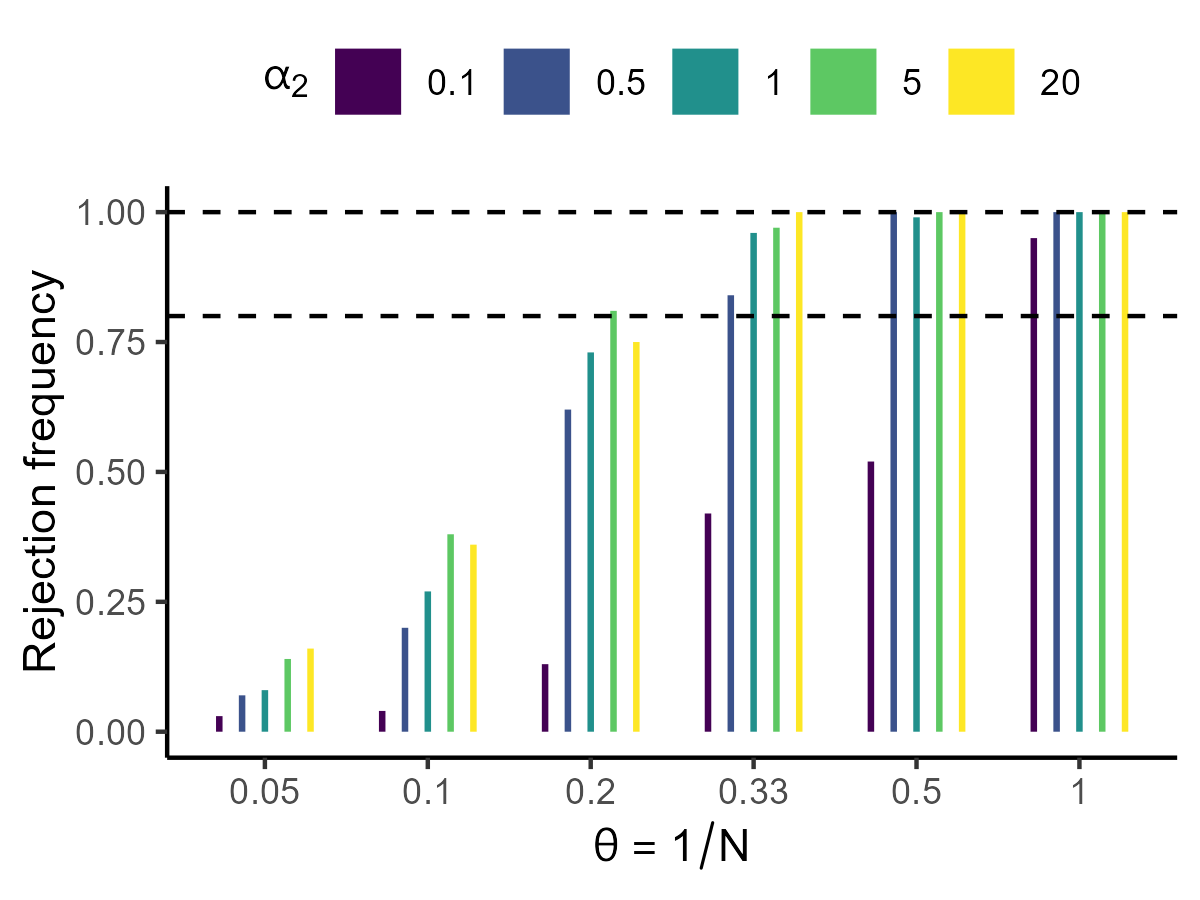}}\\
  \subfloat[$T=2000$]{\includegraphics[width = 0.32\textwidth]
  {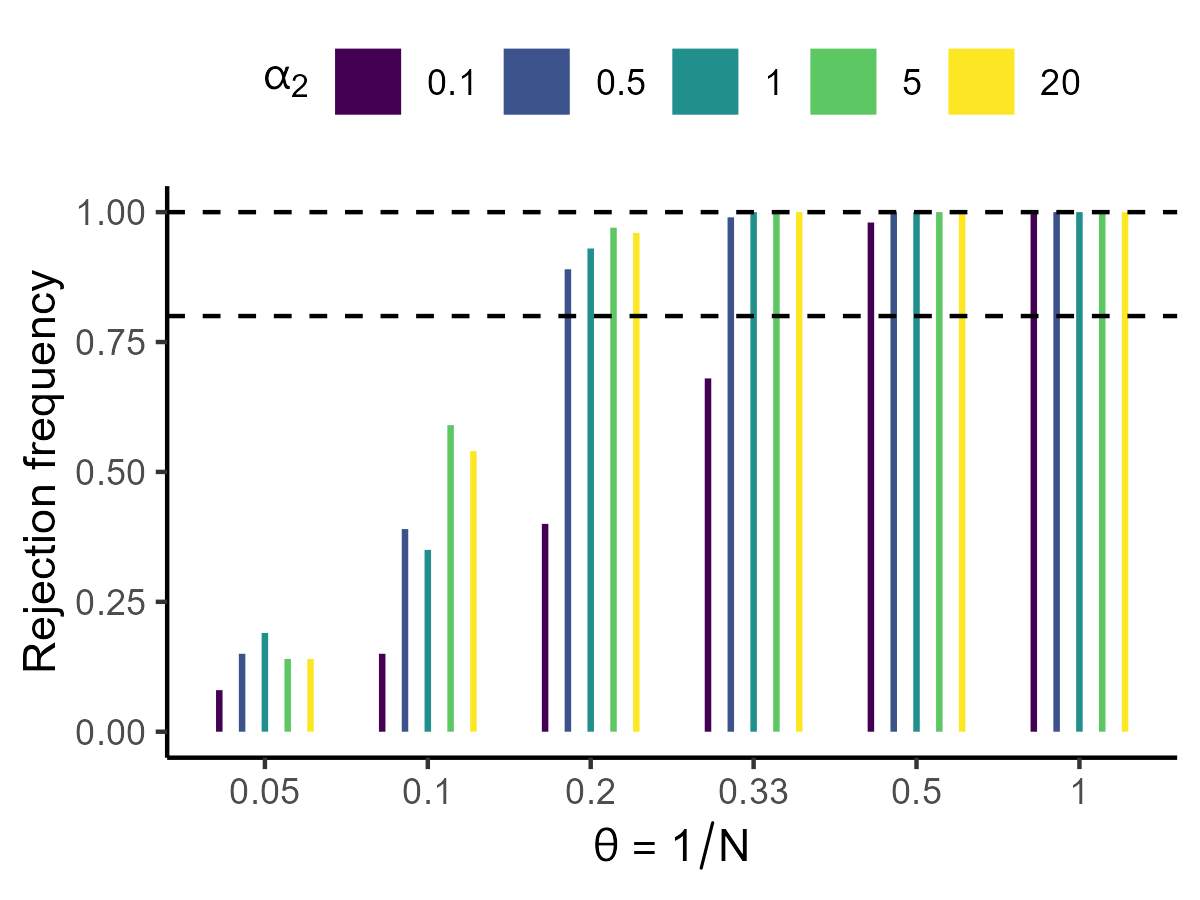}}
  \subfloat[$T=5000$]{\includegraphics[width = 0.32\textwidth]
  {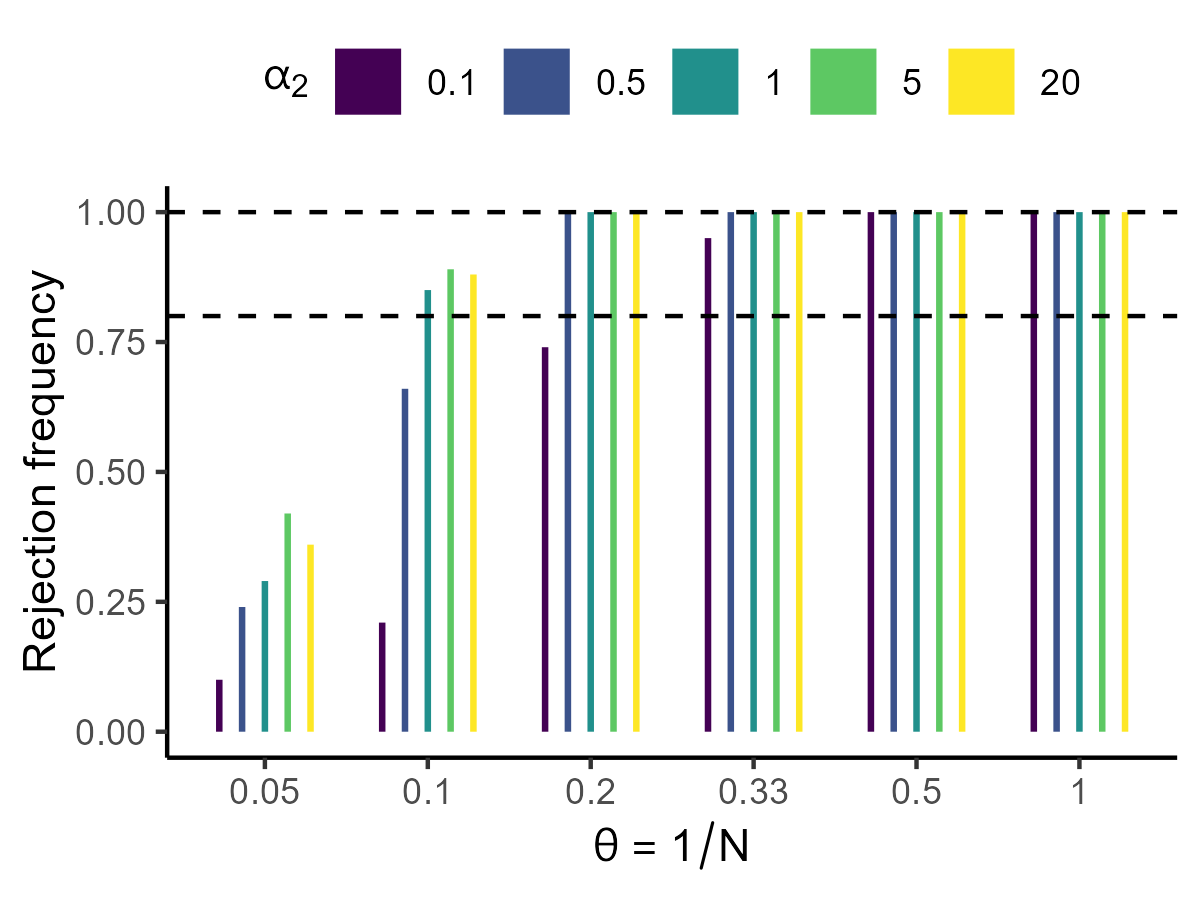}}
  \subfloat[$T=10000$]{\includegraphics[width = 0.32\textwidth]
  {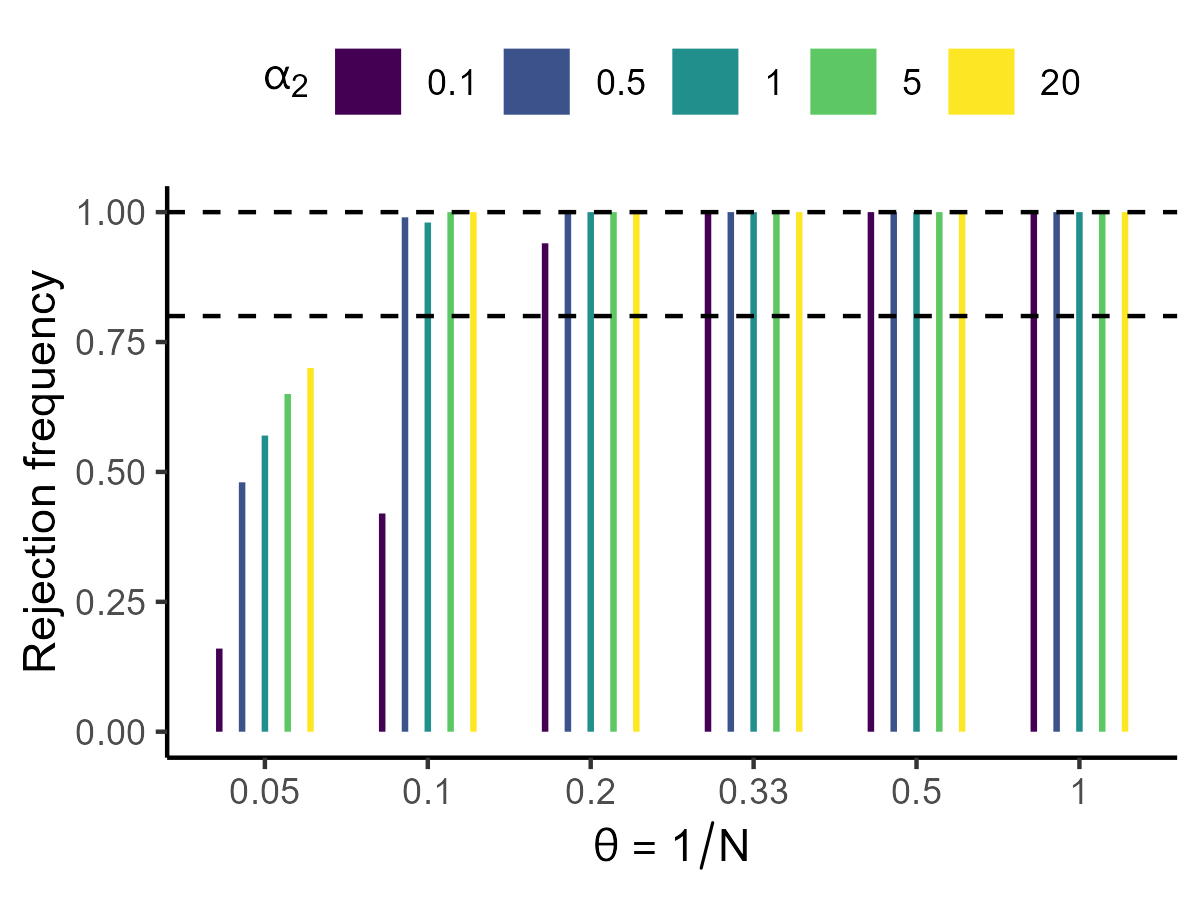}}
  \caption{Statistical power of conduct parameter $\theta$}
  \label{fg:theta_hat_power}
  \end{center}
  \footnotesize
  Note: Dotted lines are 80\% and 100\% rejection frequencies out of 100 simulation data.
\end{figure} 

\begin{figure}[!ht]
  \begin{center}
  \subfloat[$T=100$]{\includegraphics[width = 0.32\textwidth]
  {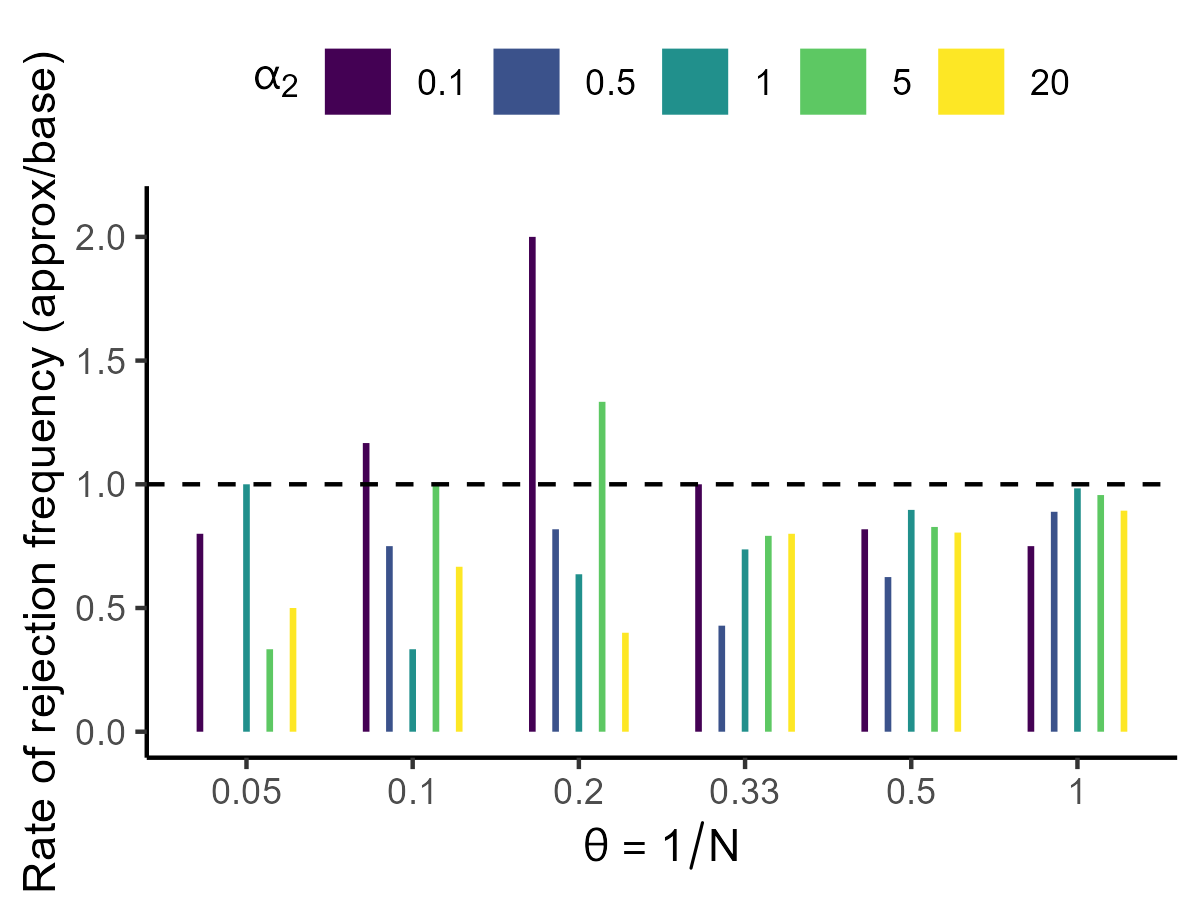}}
  \subfloat[$T=200$]{\includegraphics[width = 0.32\textwidth]
  {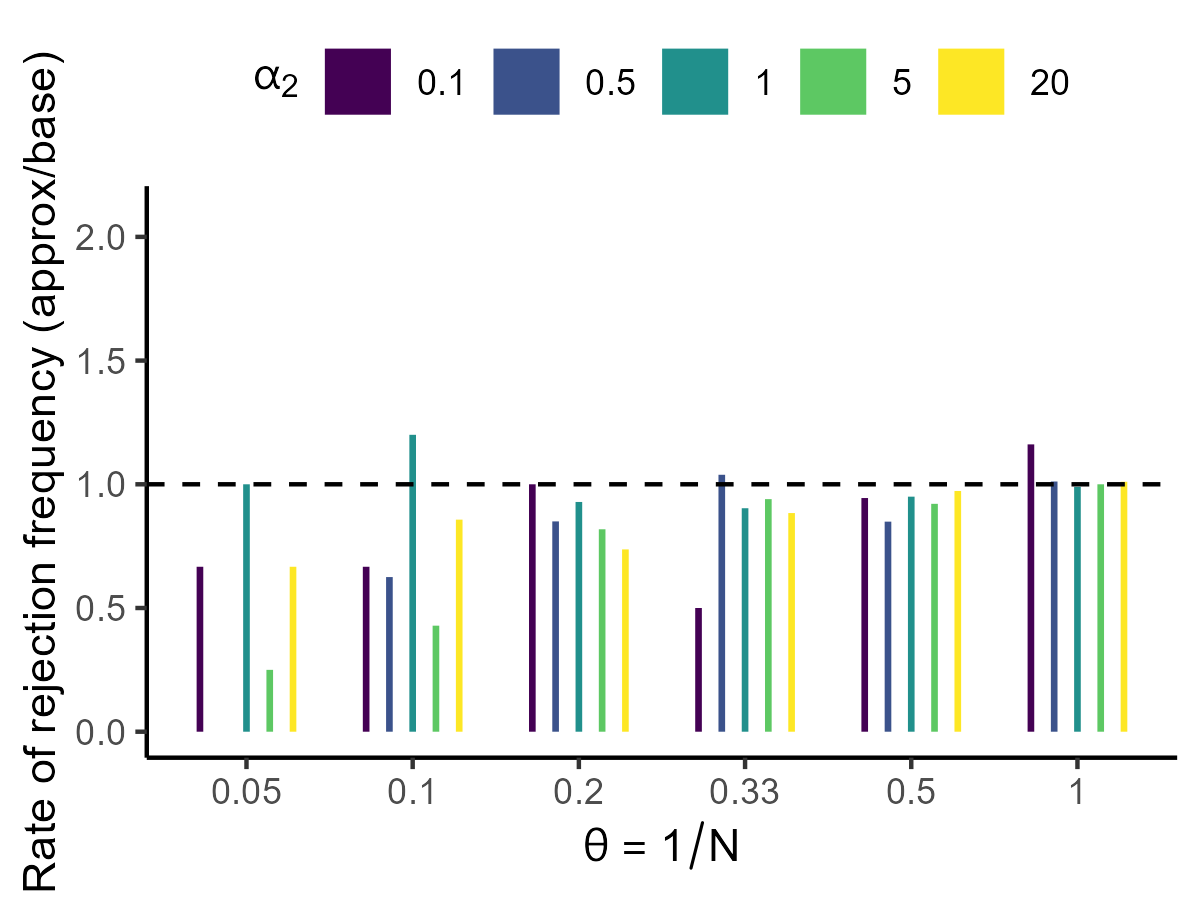}}
  \subfloat[$T=1000$]{\includegraphics[width = 0.32\textwidth]
  {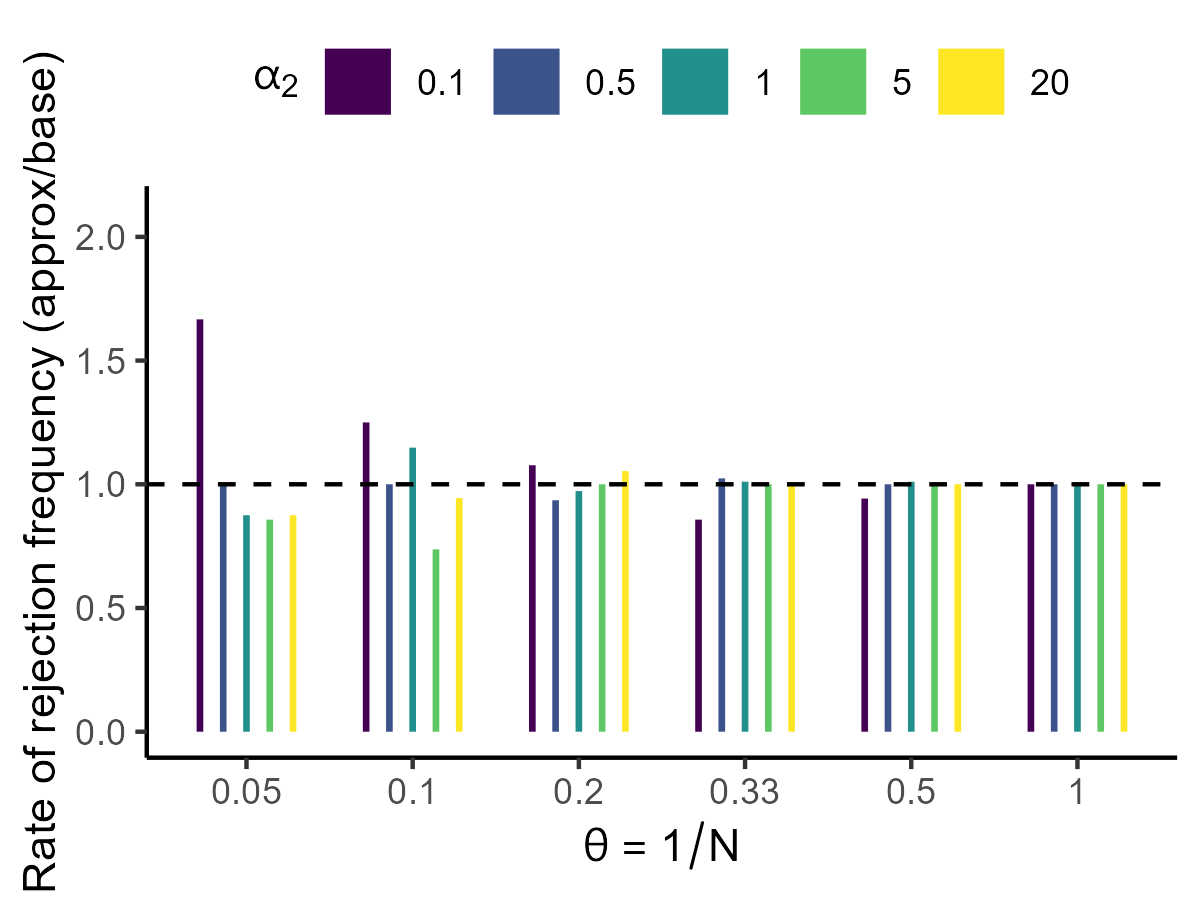}}\\
  \subfloat[$T=2000$]{\includegraphics[width = 0.32\textwidth]
  {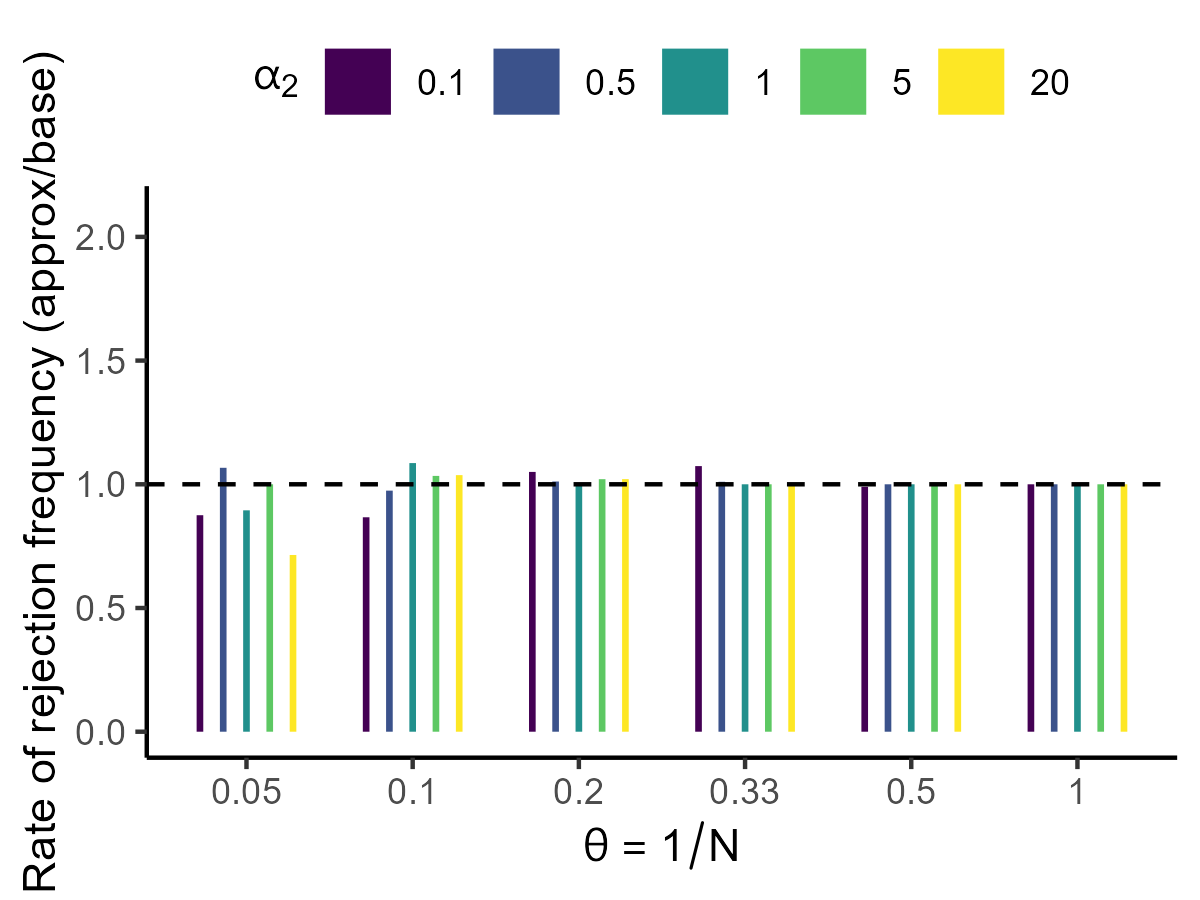}}
  \subfloat[$T=5000$]{\includegraphics[width = 0.32\textwidth]
  {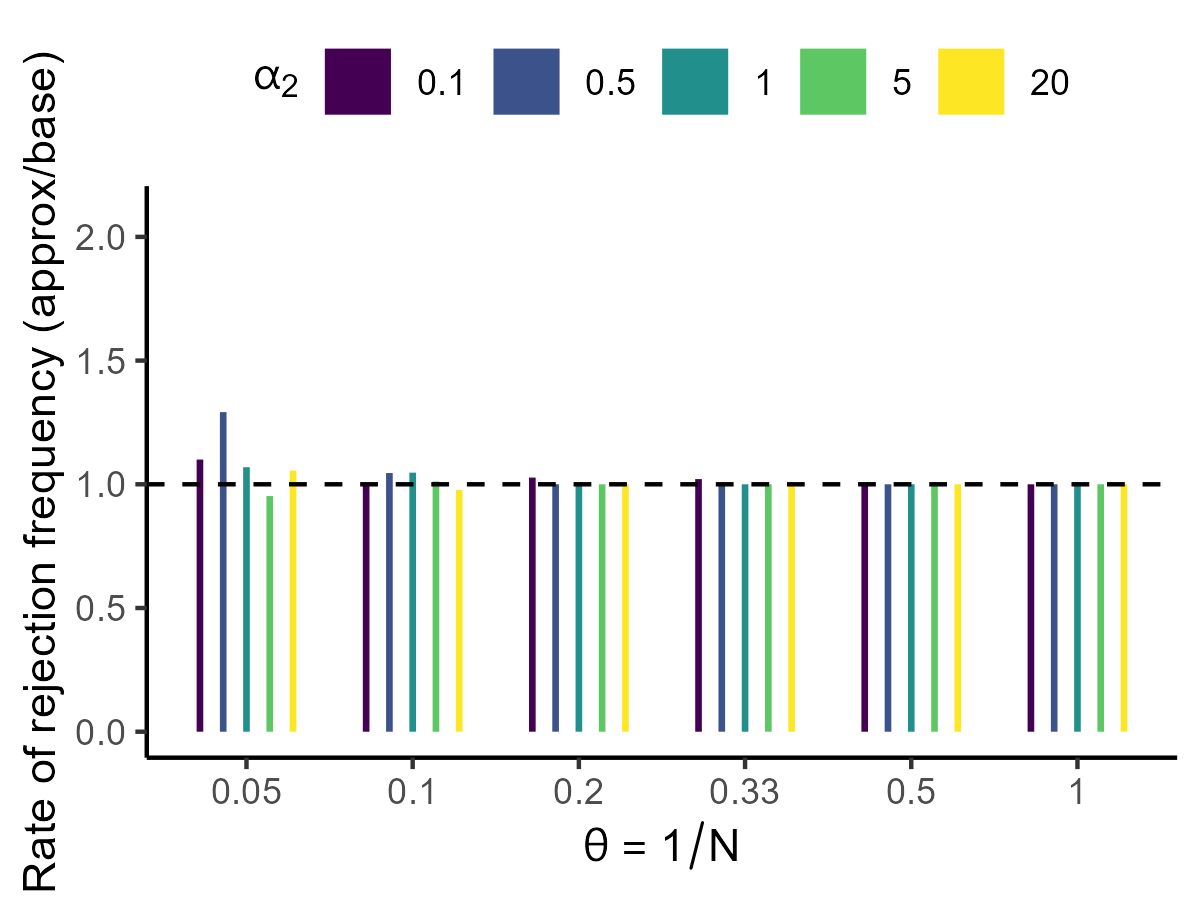}}
  \subfloat[$T=10000$]{\includegraphics[width = 0.32\textwidth]
  {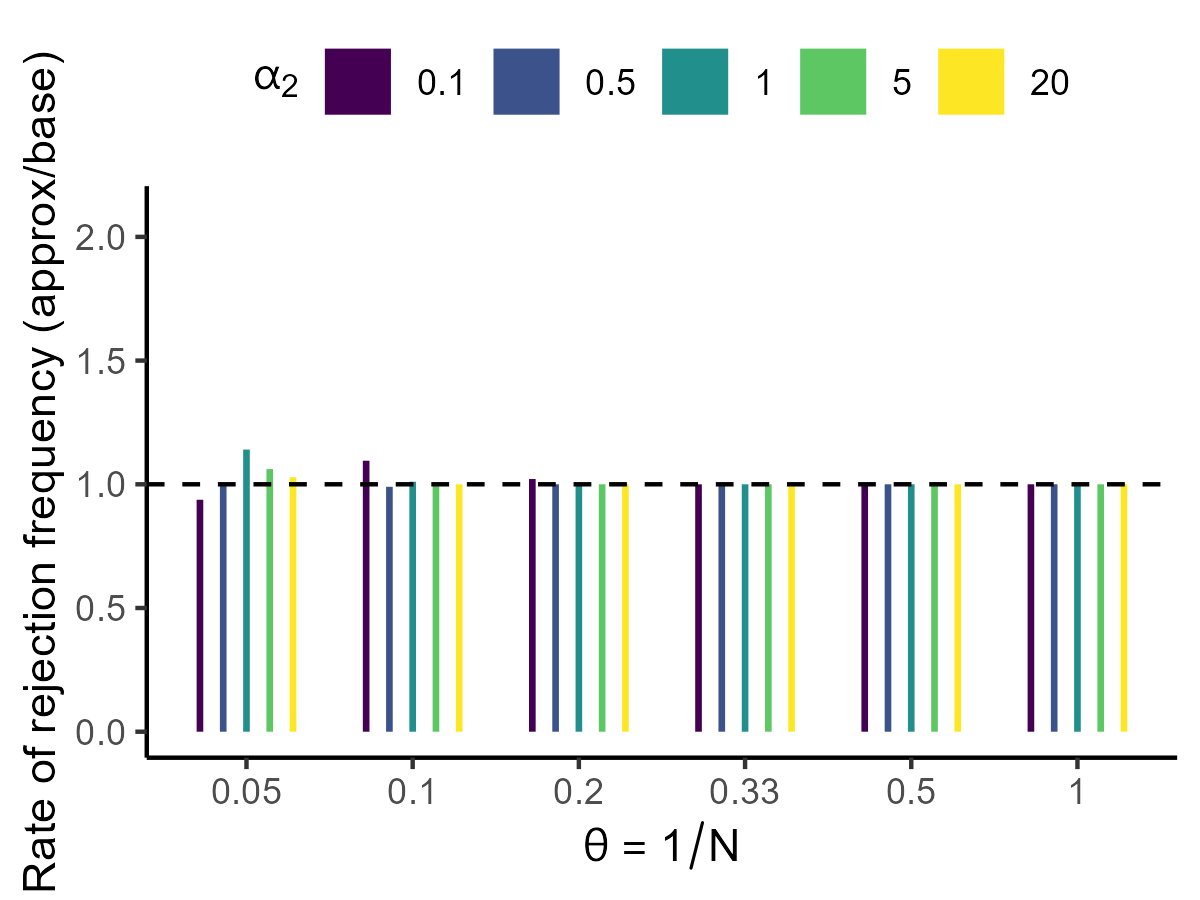}}
  \caption{Relative efficiency gain of optimal instruments in the first approach}
  \label{fg:theta_hat_power_iv_polynomial_relative_to_benchmark}
  \end{center}
  \footnotesize
  Note: Dotted lines are 80\% and 100\% rejection frequencies out of 100 simulation data.
\end{figure} 

Figure \ref{fg:theta_hat_power} presents the finite sample performance results for the conduct parameter $\theta$.\footnote{Simulation details and additional results for all other parameters are available in the online appendix.} The rejection frequency increases under the following conditions: a larger sample size (number of markets), a higher $\theta$ (fewer firms), and a larger $\alpha_2$ (stronger demand rotation instrument). Panel (f) shows that with 20 symmetric firms ($\theta=0.05$) and a sufficiently large number of markets, the power to reject the null hypothesis of perfect competition reaches approximately 70\%. However, under 20-firm symmetric Cournot competition, the null hypothesis cannot be rejected with an acceptable sample size and power.

A noteworthy finding is that even with a moderate number of markets (e.g., 1,000 in Panel (c)) and five firms, the rejection frequency cannot achieve 80\% (i.e., $1-\beta=0.8$, where $\beta$ represents the probability of a Type II error), regardless of instrument strength. This suggests that studies such as \cite{genesove1998testing} with 97 markets, \cite{shaffer1993test} with 25 markets, and \cite{steen1999testing} with 48 markets fail to reject perfect competition primarily due to small sample sizes.

\subsection{Optimal instruments results}
\begin{figure}[!ht]
  \begin{center}
  \subfloat[$T=100$]{\includegraphics[width = 0.32\textwidth]
  {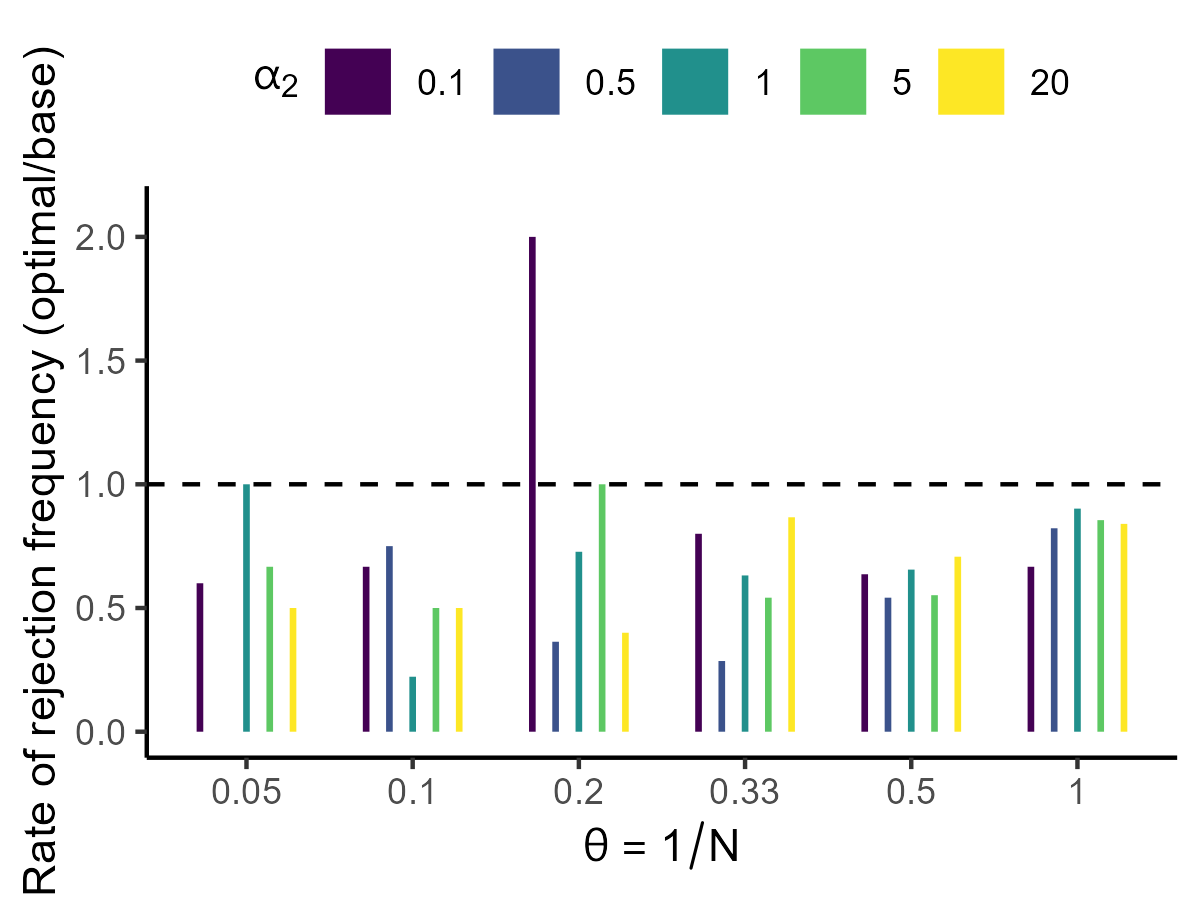}}
  \subfloat[$T=200$]{\includegraphics[width = 0.32\textwidth]
  {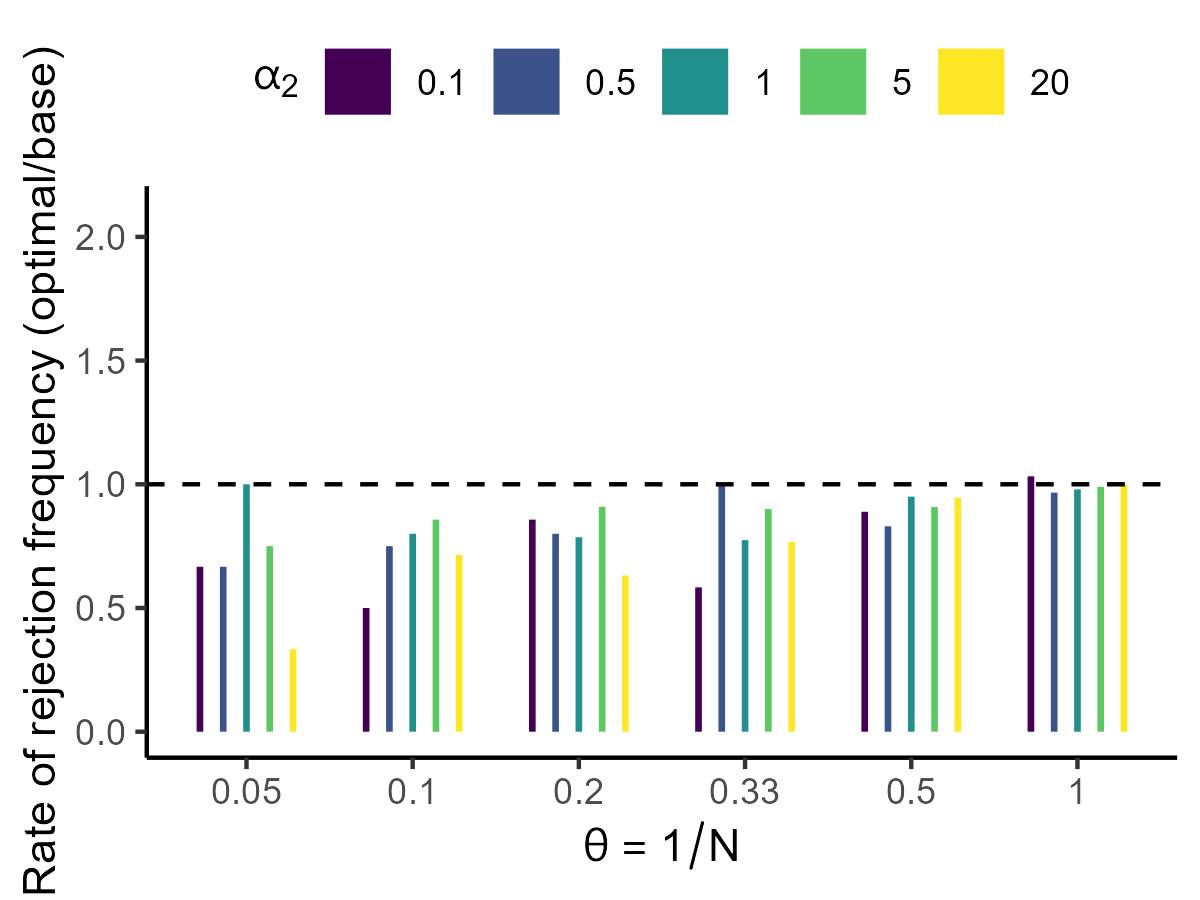}}
  \subfloat[$T=1000$]{\includegraphics[width = 0.32\textwidth]
  {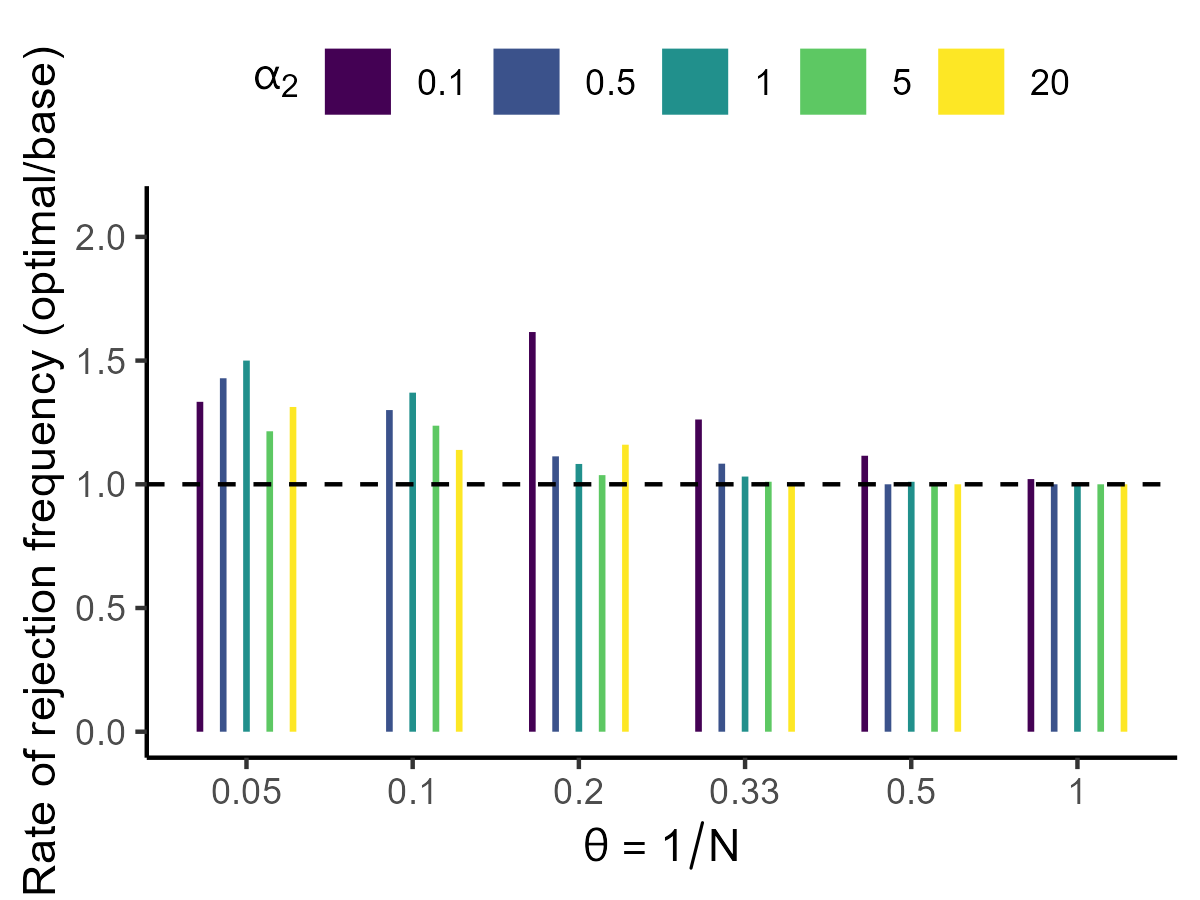}}\\
  \subfloat[$T=2000$]{\includegraphics[width = 0.32\textwidth]
  {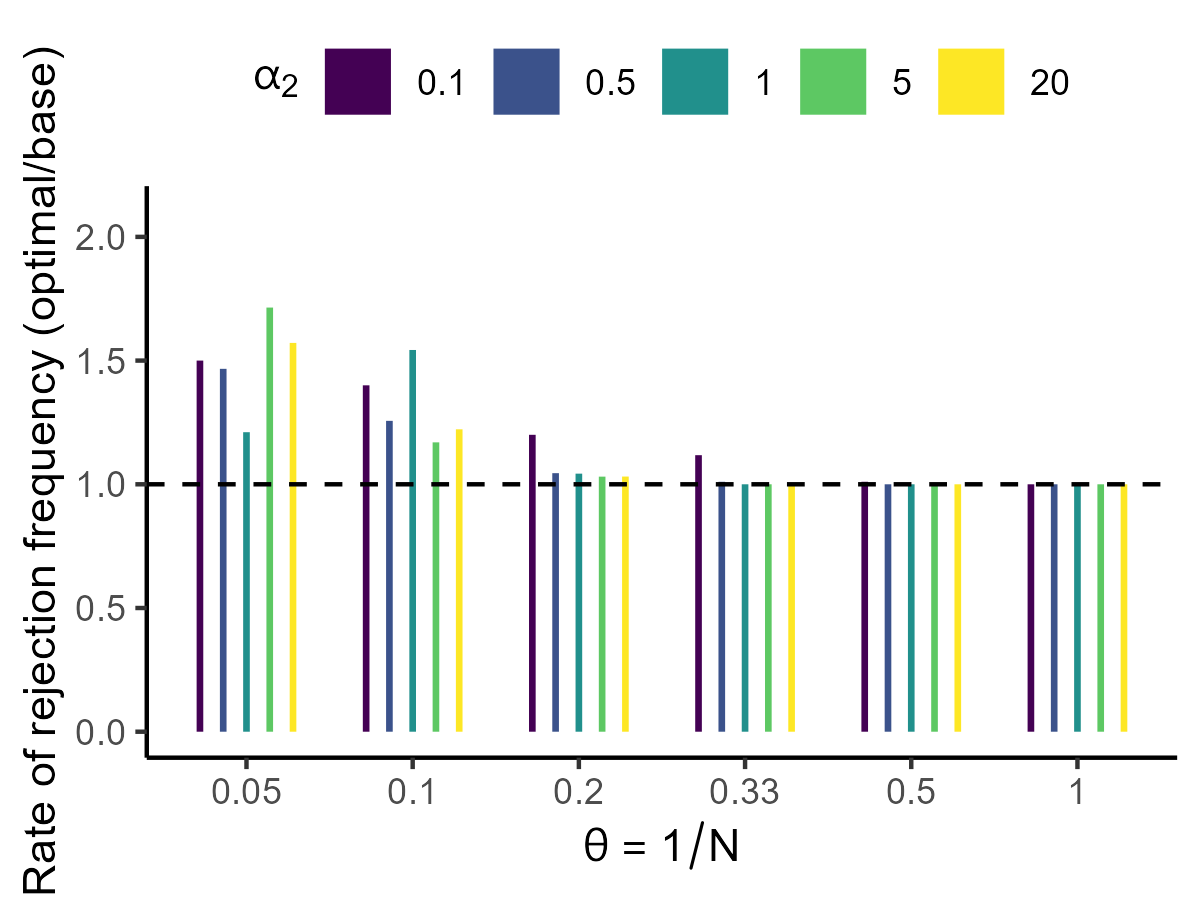}}
  \subfloat[$T=5000$]{\includegraphics[width = 0.32\textwidth]
  {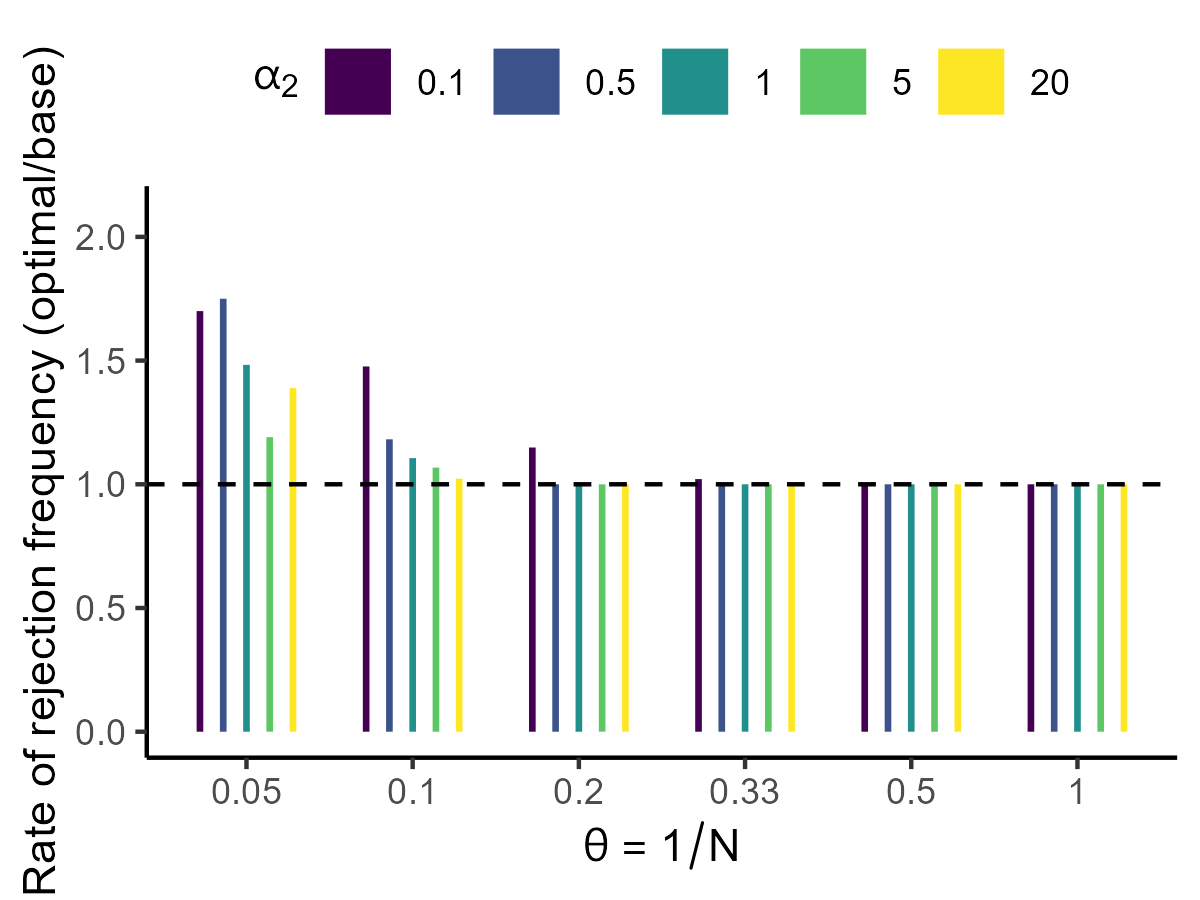}}
  \subfloat[$T=10000$]{\includegraphics[width = 0.32\textwidth]
  {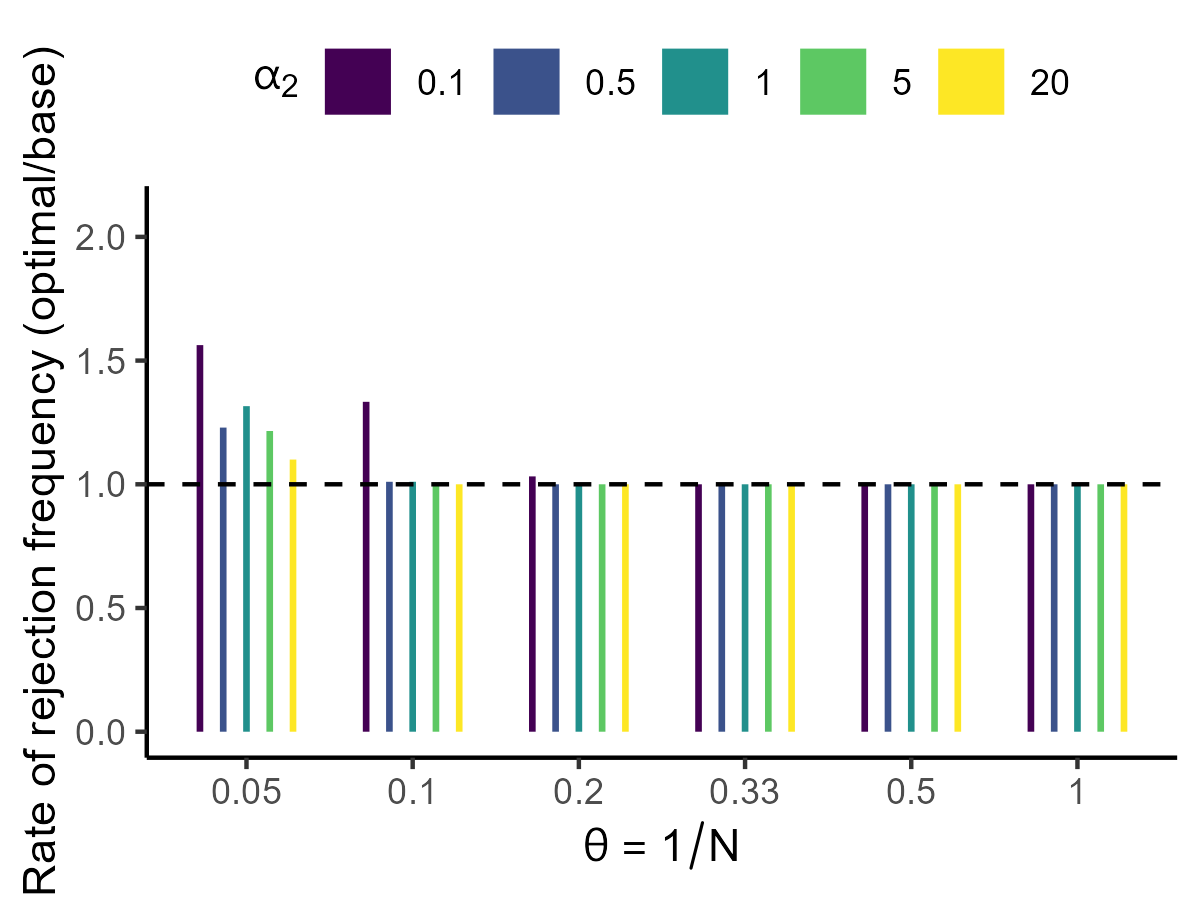}}
  \caption{Relative efficiency gain of optimal instruments in the second approach}
  \label{fg:theta_hat_power_iv_optimal_relative_to_benchmark_10000}
  \end{center}
\end{figure}

The case with optimal instruments represents the ideal scenario in which researchers can obtain the most efficient moment conditions. Figures \ref{fg:theta_hat_power_iv_polynomial_relative_to_benchmark} and \ref{fg:theta_hat_power_iv_optimal_relative_to_benchmark_10000} present the relative efficiency gains of the first and second approximation approaches compared to the benchmark model. Unfortunately, even in this best-case scenario, the efficiency gain is negligible. While optimal instruments increase the rejection probability when the number of markets exceeds 1,000, this gain does not substantially alter our benchmark results.

Why is it statistically challenging to differentiate between perfect and Cournot competition?
In differential product markets, as demonstrated by \citet{berry2014identification}, the variation in instrumental variables can aid in discerning firm behavior. 
Various factors such as changes in the number of products, prices in other markets, and alterations in product characteristics, can be utilized without requiring a specific functional form.
In contrast, homogeneous product markets exhibit limited variation only on demand rotation instruments. 
Therefore, even when the number of markets is substantial, firm conduct tests may lack the necessary power to differentiate between perfect and Cournot competition.

\section{Conclusion}
We theoretically prove why statistically rejecting the null hypothesis of Cournot and perfect competition is challenging, known as a common problem in the literature. 
We perform a statistical power analysis for conduct parameter estimation. Power rises with an increase in the number of markets, a larger conduct parameter, and a stronger demand rotation instrument. 
Nevertheless, rejecting the null hypothesis of markets operating under perfect competition remains challenging, even with a moderate number of markets (e.g., 1000) and five firms, regardless of instrument strength and the use of optimal instruments. 
This reaffirms that the difficulty in testing perfect competition, as observed by \cite{genesove1998testing}, \cite{steen1999testing}, and \cite{shaffer1993test}, is primarily attributed to the limited number of markets, rather than methodological shortcomings.

\bibliographystyle{aer}
\bibliography{conduct_parameter}

\newpage

\setcounter{page}{1}
\appendix
\section{Online appendix}\label{sec:appendix}

\subsection{Details for our simulation settings}

To generate the simulation data, for each model, we first generate the exogenous variables $Y_{t}, Z^{R}_{t}, W_{t}, R_{t}, H_{t}$, and $K_{t}$ and the error terms $\varepsilon_{t}^c$ and $\varepsilon_{t}^d$ based on the data generation process in Table \ref{tb:parameter_setting}.
We compute the equilibrium quantity $Q_{t}$ for the linear model by \eqref{eq:quantity_linear}.
We then compute the equilibrium price $P_{t}$ by substituting $Q_{t}$ and other variables into the inverse demand function \eqref{eq:linear_demand}.

We estimate the equations using the \texttt{ivreg} package in \texttt{R}.
An important feature of the model is that we have an interaction term of the endogenous variable $Q_{t}$ and the instrumental variable $Z^{R}_{t}$.
The \texttt{ivreg} package automatically detects that the endogenous variables are $Q_{t}$ and the interaction term $Z^{R}_{t}Q_{t}$, running the first stage regression for each endogenous variable with the same instruments. To confirm this, we manually write R code to implement the 2SLS model. 
When the first stage includes only the regression of $Q_{t}$, estimation results from our code differ from the results from \texttt{ivreg}. 
However, when we modify the code to regress $Z^{R}_{t}Q_{t}$ on the instrument variables and estimate the second stage by using the predicted values of $Q_{t}$ and $Z^{R}_{t}Q_{t}$, the result from our code and the result from \texttt{ivreg} coincide.



\subsection{Additional results}
Estimation results of all parameters except conduct parameter $\theta$ are shown on the author's github. 
Totally, we confirm that estimation is accurate with small bias and root-mean-squared error when increasing sample size, as in \cite{matsumura2023resolving}.
Interested readers can modify the current data-generating process.

\end{document}